\documentclass[twocolumn,twoside,slac]{revtex4}
\usepackage{graphicx}
\usepackage{fancyhdr}
\begin{document}

\title{The PROOF Distributed Parallel Analysis Framework based on ROOT}


\author{Maarten Ballintijn}
\affiliation{MIT, Cambridge}
\author{Rene Brun}
\affiliation{CERN, ROOT Project}
\author{Fons Rademakers}
\affiliation{CERN, ALICE Collaboration and ROOT Project}
\author{Gunther Roland}
\affiliation{MIT, Cambridge}

%%%%%%%%%%%%%%%%%%%%%%%%%%%%%%%%%%%%%%%%%%%%%%%%%%%%%%%%%%%%%%%%%%%%%%%%%
\begin{abstract}
%%%%%%%%%%%%%%%%%%%%%%%%%%%%%%%%%%%%%%%%%%%%%%%%%%%%%%%%%%%%%%%%%%%%%%%%%

The development of the Parallel ROOT Facility, PROOF, enables a
physicist to analyze and understand much larger data sets on a shorter
time scale. It makes use of the inherent parallelism in event data and
implements an architecture that optimizes I/O and CPU utilization in
heterogeneous clusters with distributed storage. The system provides
transparent and interactive access to gigabytes today. Being part of
the ROOT framework PROOF inherits the benefits of a performant object
storage system and a wealth of statistical and visualization tools.
This paper describes the key principles of the PROOF
architecture and the implementation of the system. We will illustrate its
features using a simple example and present measurements of the
scalability of the system. Finally we will discuss how PROOF can be
interfaced and make use of the different Grid solutions.

%%%%%%%%%%%%%%%%%%%%%%%%%%%%%%%%%%%%%%%%%%%%%%%%%%%%%%%%%%%%%%%%%%%%%%%%%
\end{abstract}
%%%%%%%%%%%%%%%%%%%%%%%%%%%%%%%%%%%%%%%%%%%%%%%%%%%%%%%%%%%%%%%%%%%%%%%%%

\maketitle

\thispagestyle{fancy}


%%%%%%%%%%%%%%%%%%%%%%%%%%%%%%%%%%%%%%%%%%%%%%%%%%%%%%%%%%%%%%%%%%%%%%%%%

\section{INTRODUCTION}

The Parallel ROOT Facility, PROOF, is an extension of the well known
ROOT system (http://root.cern.ch/)~\cite{ROOT96} that allows the easy
and transparent analysis of large sets of ROOT files in parallel on
remote computer clusters. The main design goals for the PROOF system
are transparency, scalability and adaptability. With transparency we
mean that there should be as little difference as possible between a local
ROOT based analysis session and a remote parallel PROOF session, both
being interactive and giving the same results.
With scalability we mean that the basic architecture should not
put any implicit limitations on the number of computers that can be
used in parallel. And with adaptability we mean that the system should
be able to adapt itself to variations in the remote environment (changing
load on the cluster nodes, network interruptions, etc.).

Being an extension of the ROOT system, PROOF is designed to work on
objects in ROOT data stores. These objects can be individually
{\em keyed\/} objects as well as {\tt TTree} based object collections.
By logically grouping many ROOT files into a single object very large
data sets can be created. In a local cluster environment these data
files can be distributed over the disks of the cluster nodes or
made available via a NAS or SAN solution.

In the near future, by employing Grid technologies, we plan to extend
PROOF from single clusters to virtual global clusters. In such an
environment the processing may take longer (not interactive), but the
user will still be presented with a single result, like the processing
was done locally.

The PROOF development is a joint effort between CERN and MIT.

%%%%%%%%%%%%%%%%%%%%%%%%%%%%%%%%%%%%%%%%%%%%%%%%%%%%%%%%%%%%%%%%%%%%%%%%%
\section{THE PROOF SYSTEM}

\subsection{System Architecture}

\begin{figure}
\includegraphics[width=65mm]{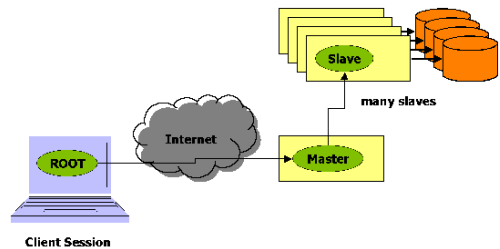}
\caption{PROOF Architecture overview}
\label{proof_simple}
\end{figure}

PROOF consists of a 3-tier architecture, the ROOT client session,
the PROOF master server and the PROOF slave servers. The user connects
from his ROOT session to a master server on a remote cluster and the
master server in turn creates slave servers on all the nodes in the
cluster. Queries are processed in parallel by all the slave servers.
Using a pull protocol the slave servers ask the master for work packets,
which allows the master to distribute customized packets for each slave
server. Slower slaves get smaller work packets than faster ones and faster
ones process more packets. In this scheme the parallel processing performance
is a function of the duration of each small job, packet, and the networking
bandwidth and latency. Since the bandwidth and latency of a networked
cluster are fixed the main tunable parameter in this scheme is the
packet size. If the packet size is chosen too small the parallelism
will suffer due to the communication overhead caused by the many
packets sent over the network between the master and the slave servers.
If the packet size is too large the effect of the difference in performance
of each node is not evened out sufficiently. This allows the PROOF system
to adapt itself to the performance and load on each individual cluster node
and to optimize the job execution time.

%- Goals: transparency, scalability, adaptability
%  (explained in detail below)
%
%- analyze Trees, Objects, parallel execution of scripts
%
%- forward ref TSelector, TDSets and PAR
%
%- Data access strategy
%   - Allocation of CPU resources
%   - Local data first, rootd/rfio, SAN/NAS
%
%- Input objects
%
%- Output objects, merge paradigm
%
%- pull architectutre
%
%- sandbox and cache features
%
%- authentication
%
%- CINT vs. compiled

\subsection{{\tt TSelector} Framework}

The {\tt TSelector} framework plays a central role in the PROOF
system. It allows users to write their analysis code in a way
that enables them to process small data samples locally on
a workstation and using the same code to process large data sets
in parallel on a cluster or the grid using PROOF.

\begin{table}
\begin{verbatim}
  class TSelector {
    TList      *fInput;
    TList      *fOutput;
    TObject    *fObject;

    void        Init(TTree *tree);
    void        Begin(TTree *tree);
    const char *GetOption() const;
    Bool_t      Process(Int_t entry);
    void        Terminate();
  };
\end{verbatim}
\caption{Simplified {\tt TSelector} class definition\label{selector_api}}
\end{table}

To use the framework a user derives a class from {\tt TSelector}
and implements the member functions that are part of the protocol.
The framework also specifies how input objects are
made available to the selector object and
how output objects are returned to the user or client session.

The {\tt TTree::MakeSelector()} function generates a
skeleton version of a {\tt TSelector}
derived class that should be used as the basis for the
analysis code. The user can either edit  the generated
files directly or create a class that inherits from the
generated class. The second option makes sense if the
definition of the tree is expected to change. It allows
for the class to be easily regenerated without modifying the
user class.
Table~\ref{selector_api} lists the most important functions
and members which we will describe in some more detail.

The input objects are made available to the selector
in the list pointed to by {\tt fInput}. The list is available
throughout the life of the selector. The objects
in the list are owned by the system. Each of the slaves
has a full copy of the list.

The {\tt Begin()} function is called before any object or event is
processed. It is a good place to create things like
histograms and initialize data that is needed to
process each event. The {\tt Begin()} function is called
once in each slave.

The {\tt Init()} function is called each time a new file is
opened by the system. The generated code takes care
of setting up the branches in case of a tree analysis.
The user can extend this routine for example to read a
calibration object that pertains to the objects or tree in that file.

The {\tt Process()} function is called for each object or event in
the tree. When processing a tree it is up to the
user to decided which branches should be read and when.
Generally it is most efficient to read branches as they
are used. E.g. read the branch(es) that are used to make
the selection and only if the event is to be processed read
the rest of the branches that are needed. When processing
{\em keyed} objects the pointer to the object is stored in the
{\tt fObject} member.

The {\tt Terminate()} function is called after all objects or events have been
processed. It allows the user to do a cleanup, like deleting temporary
objects and closing auxiliary files. It is also the place to do final
calculations. The {\tt Terminate()} function is called once in each slave.

Finally the contents of the {\tt fOutput}
lists of all the slaves are merged in the master and send back to the client.
The merging uses the special {\tt Merge(Tlist*)} API. The objects are
fist grouped by name.
Using the introspection features provided by CINT~\cite{CINT}, the C++
interpreter used by ROOT, it is then determined
if the {\tt Merge()} function is implemented for those objects.
For histograms
this is available by default. User classes can implement the API if merging
makes sense. If the merge function is not available the individual objects
are returned to the client.

\subsection{Specifying Data Sets}

The {\tt TDSet} class is used to specify the collection of objects or
trees that are to be processed with PROOF. When analyzing trees
the user must pass {\tt "TTree"} to the TDSet constructor. The
{\tt TDSet::Add()} function
is then used to add a file, directory within that file and tree name
to the set.

To process a collection of objects, the user must pass the
name of the class of the objects to the constructor. In that case
{\tt TDSet::Add()}
is used to add files and the directories within them that contain
the objects.

The contents of one {\tt TDSet} can also be added to another {\tt TDSet}.
The {\tt TDSet} implements the {\tt Print()} function which allows the contents
to be inspected. Users can create {\tt TDSets} "by hand" but it is also
foreseen that (Grid) catalogs will provide query interfaces that
return {\tt TDSets}. It will be possible to use logical filenames rather then
physical filenames to specify files. The translation will be done using the
API defined by the abstract class {\tt TGrid} in combination with
plugins implementing the communication with available catalogs.

\subsection{The PROOF Package Manager}

In complex analysis environments the analysis scripts very likely depend
on one or more external libraries containing common algorithms. To be
able to run these scripts successfully on PROOF it is required that these
external libraries are available on each slave node.

The PROOF package manager has been designed to distribute and install
these libraries on a PROOF cluster. Packages are compressed {\em tar}
files containing the library source or binaries with one additional
{\tt PROOF-INF} directory (like Java JAR files).
This {\tt PROOF-INF} directory contains a {\tt SETUP.C} script and,
optionally, a {\tt BUILD.sh} shell script. The {\tt BUILD.sh} script is
used to compile the library source on the PROOF cluster, which might have a
different type of CPU or OS than the user's local system. The {\tt SETUP.C}
script is used to load the libraries into the PROOF server processes.
Package files are called PAR (PROOF ARchive) files and must have a
{\tt .par} extension.

For example an {\tt Event.par} file which provides a library called
{\tt libEvent.so} has the following {\tt BUILD.sh}:
\begin{verbatim}
  make libEvent.so
\end{verbatim}
and {\tt SETUP.C}:
\begin{verbatim}
  Int_t SETUP()
  {
    gSystem->Load("libEvent");
    return 1;
  }
\end{verbatim}

In a two step process packages are first uploaded
to PROOF and then enabled as required. The PROOF package API
(see table~\ref{package_api}) gives the user full control.
To avoid unnecessary transfer of par files the upload command
first asks the MD5 checksum of the remote version of the package
and, if it exists, checks it with the one of the local version.
Only when the checksums are different will the par file be transfered
to the PROOF cluster.

\begin{table}
\caption{The PROOF Package API\label{package_api}}
\begin{verbatim}
  Int_t  UploadPackage(const char *par, Int_t par = 1);
  void   ClearPackage(const char *package);
  void   ClearPackages();
  Int_t  EnablePackage(const char *package);
  void   ShowPackages(Bool_t all = kFALSE);
  void   ShowEnabledPackages(Bool_t all = kFALSE);
\end{verbatim}
\end{table}

%%%%%%%%%%%%%%%%%%%%%%%%%%%%%%%%%%%%%%%%%%%%%%%%%%%%%%%%%%%%%%%%%%%%%%%%%
\section{A REAL LIFE EXAMPLE}

In this section we will use a simple but full fledged example to
illustrate the use of PROOF and to describe some of the features
of the implementation.

\begin{figure}
\includegraphics[width=55mm]{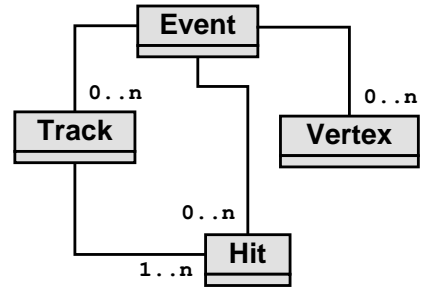}
\caption{The AnT data model}
\label{ant_datamodel}
\end{figure}

\subsection{Local Development}

The data model underlying the tree in this example is typical for a HEP
experiment (see fig~\ref{ant_datamodel}). The tree contains a
{\tt TClonesArray}
for a hit, track and vertex class and a single event class. Relations between
hits, tracks and vertexes are expressed using {\tt TRef} and {\tt TRefArray}.

For this example we use a very simple example script. The relevant parts are
included below. All error checking has been omitted for brevity.
In the {\tt Init()} function a single histogram
is created which is stored in the data member {\tt fVtx\_x} of the selector
class.

\begin{verbatim}
  void antsel::Begin(TTree *tree)
  {
    // Initialize the tree branches.
    Init(tree);

    fVtx_x = new TH1F("v","v",100,-10.,10.);
  }
\end{verbatim}

The {\tt Process()} function reads the data, makes a simple selection
and fills the histogram. It uses the {\tt fRMSSelVtx} reference to get
the proper vertex. In a more elaborate version it would only
read the event branch for the test and read the vertex branch as needed.

\begin{verbatim}
  void antsel::Process(Int_t entry)
  {
    fChain->GetTree()->GetEntry(entry);

    if (eventInfo->fPdlMean > 1500) {
      TPhAnTVertex *vtx = (TPhAnTVertex*)
        eventInfo->fRMSSelVtx->GetObject();
      fVtx_x->Fill(vtx->fPos.X());
    }
  }
\end{verbatim}

The {\tt Terminate()} function adds the histogram to the output list
{\tt fOutput} such that it will be returned to the client by PROOF.

\begin{verbatim}
  void antsel::Terminate()
  {
    fOutput->Add(fVtx_x);
  }
\end{verbatim}

Running the script on a typical laptop we can analyze
2000 events, or 8~Mbyte, in about 6~seconds (ROOT I/O
has compressed the data in the tree by a factor 5.)
Once the script works as required we are ready to run it using
PROOF.

\subsection{Session Setup}

We start by creating a PROOF session. The argument specifies the PROOF
cluster we want to connect to.

\begin{verbatim}
  root[1] gROOT->Proof("pgate.lns.mit.edu")
\end{verbatim}

We first have to authenticate ourselves to the PROOF server. All
authentication methods implemented in the standard ROOT {\tt TAuthenticate}
class are available.
The PROOF master server is then created. The master server reads the
{\tt proof.conf} configuration file and uses the information to start
a number of slave servers.
The configuration file is provided by the administrator of the PROOF cluster.
It contains information about the nodes that make up the PROOF cluster,
the relative performance of the nodes and which nodes share a file system.

When all the slaves are started we are ready to run queries.
But before that we can configure packages. For this example we'll upload
and enable a single package

\begin{verbatim}
  root[2] gProof->UploadPackage("ant.par")
  root[3] gProof->EnablePackage("ant")
\end{verbatim}

Normally we only need to upload a package if it has changed. The
{\tt EnablePackage()} function will cause the package to be build
in the case of a {\em source} package. Finally the
{\tt SETUP.C} script is run on all slaves, loading our library and
making available its classes.

We will use a previously loaded script to create the data set
\begin{verbatim}
  root[4] TDSet *d = get_dataset()
\end{verbatim}
Instead of creating the {\tt TDSet} it might be returned by a database
or catalog query.

\subsection{PROOF Run}

\begin{figure*}[t]
\centering
\includegraphics[width=135mm]{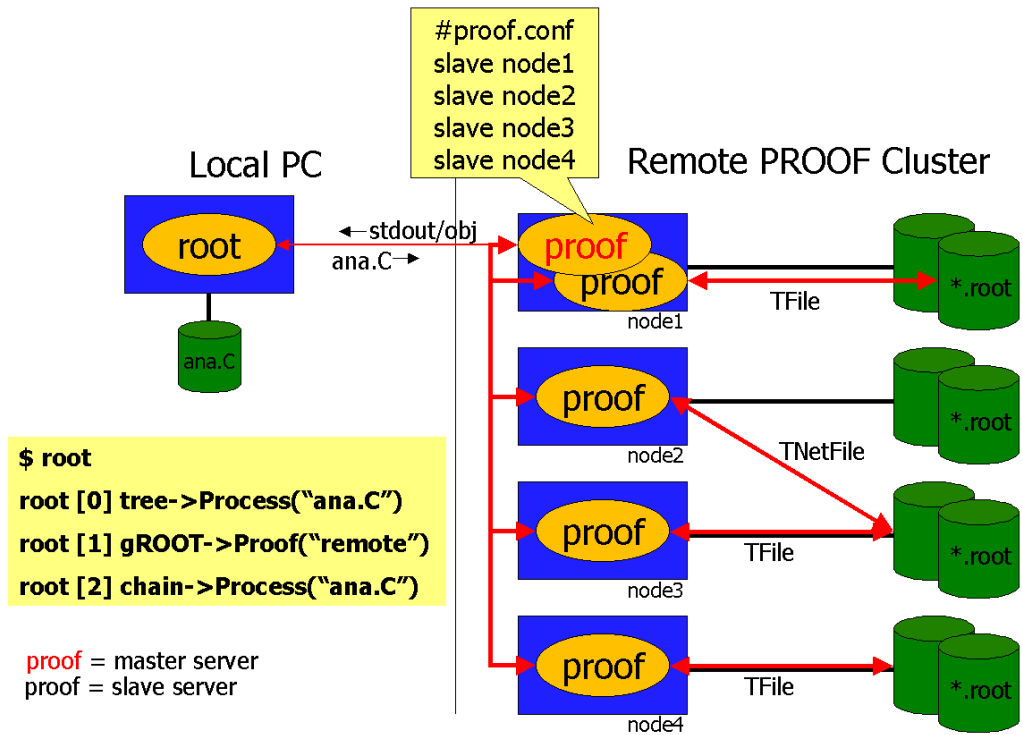}
\caption{Detailed Proof Session}\label{detailed_proof}
\end{figure*}

We are now ready to use PROOF to process our query. Using the same
script that was used previously and the data set we just created
we issue the command

\begin{verbatim}
  root[5] d->Process("antsel.C", "", 60000)
\end{verbatim}

The system goes through a number of steps to implement this command.
First the selector script and possibly its corresponding include
file are send to the master and from the master to the slaves.
The system will optimize this step if an up to date version
of the script is already available in the cluster. It also makes use
of the shared file system(s) if available.

Then the client sends the query message, including the
input objects, option and parameters, to the master.
The master determines the total number of events in the dataset
by having the slaves open the files in the dataset in parallel.
This information could in the future also be obtained from a database,
avoiding this step in he process. The master server creates a list of
nodes with the files to be processed and their sizes. This list
will be used to optimize the distribution of work over the slaves.
The files have to be specified as {\tt rootd} URLs for this optimization
to be available, e.g. {\tt "root://proof.mit.edu/data/file.root"}.

The master now forwards the query to the slaves. Each slave starts
the script and enters a loop in which it asks the master for work packets.
The master will allocate work to each slave using the information about
the location of the files. A slave will first be assigned files which
are on its local disk. When a slave has exhausted all local files
the {\tt rootd} protocol is used to process files on other nodes.
If no location information was available, .e.g if the files are stored
on a central NFS server, the algorithm reduces to a round robin assignment
of the files to the slaves. An heuristic is used to determine the packet size.
It takes into account the number of slaves and their relative performance
as well as the total number of events to be processed. The master monitors
the real-time performance of each slave allowing this heuristic to be refined.
The master also records which packet is processed by which slave allowing
error recovery to be implemented in the case of slave failure.

When all events are processed the slaves send the partial results to the master.
Using the previously described {\tt Merge()} algorithm the master combines the
partial results and sends them to the client.

\subsection{The Results}

The processing of the above query ran in about 12~seconds using 8~slaves
on 4~dual Athlon 1.4~GHz machines. The amount of data processed was 240~Mbyte.
This shows the large improvement of even a small dedicated cluster
over a typical desktop workstation. A more precise measurement of the
PROOF performance will be presented in the next section.

%TODO: should we add a top of page wide 2 graphs with low and high statistics
%like in the ppt ?

%%%%%%%%%%%%%%%%%%%%%%%%%%%%%%%%%%%%%%%%%%%%%%%%%%%%%%%%%%%%%%%%%%%%%%%%%
\section{SCALABILITY MEASUREMENTS}

\begin{figure}
\includegraphics[width=75mm]{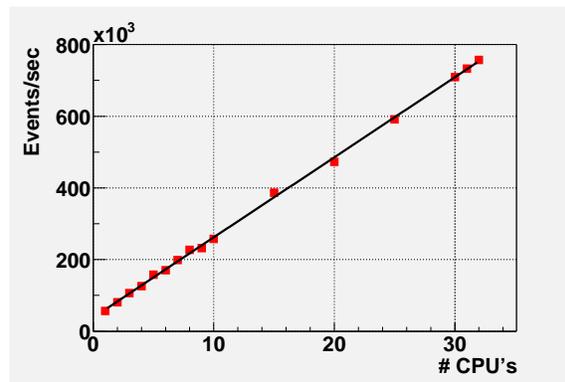}
\caption{PROOF Scalability}
\label{proof_scale}
\end{figure}

First performance measurements show a very good and efficient scalability
(see fig~\ref{proof_scale}).
For the tests we used a Linux cluster of 32 nodes. Each node had two Itanium~2
1~GHz CPU's, 2x75~GB 15K SCSI disk, 2~GB RAM and fast ethernet.
The data set to be analyzed consisted of 128 files totaling 8.8~GB of data
(9 million events). Each cluster node had 4 files (277~MB). Processing
these 128 files using one node took 325~seconds and 32~nodes in parallel
only 12~seconds. One node processing only its 4~local files took 9~seconds.
This shows that the efficiency is about 88\%. We also ran tests using
all 64~CPU's in the cluster (two slaves per node). This showed the same
linearity but less efficiency, due to overhead in the Linux SMP
implementation and resource contention in each node. All in all we
expect an efficient scalability to at least a 100~node cluster.

%%%%%%%%%%%%%%%%%%%%%%%%%%%%%%%%%%%%%%%%%%%%%%%%%%%%%%%%%%%%%%%%%%%%%%%%%
\section{PROOF IN THE GRID ENVIRONMENT}

To be able to build a global virtual PROOF cluster we need to use the
Grid services that are currently being developed.
The interfacing of PROOF to the Grid can be done at several levels.
The following levels have been identified:
\begin{itemize}
 \item Interface to the grid file catalog allowing a user to select a
  data set based on tags or logical file names (using wildcards etc).
 \item Interface to the grid resource broker to find the best location(s),
  based on the data set, where to run the query. This could trigger
  the replication of some missing files to a cluster (only when the
  amount of data in the files is relatively small).
 \item Interface to the grid job queue manager to start proof master and
  slave daemons on the remote cluster. The ROOT client will then
  connect to these pre-started daemons to create a PROOF session.
  This will require the grid queuing system to support interactive
  high priority jobs.
\end{itemize}

We are currently working with the AliEn~\cite{ALIEN} Grid developers
on a prototype that implements step wise the above scenarios.
AliEn (http://alien.cern.ch) is a grid solution tuned for typical HEP
data processing. It has been developed by the ALICE collaboration but
is experiment independent. It provides all commonly understood Grid
services, like: file catalog, replication service, job queue manager,
resource broker, authentication service, monitoring, etc. It is entirely
implemented in Perl and has a C client API. AliEn is simple to install
and works reliably. In ALICE it is routinely used for simulation and
reconstruction. AliEn has also been chosen as the grid component for
the EU MammoGrid project.

The ROOT and PROOF interface to AliEn, and other Grid middleware, will
be via a {\tt TGrid} abstract interface.

%%%%%%%%%%%%%%%%%%%%%%%%%%%%%%%%%%%%%%%%%%%%%%%%%%%%%%%%%%%%%%%%%%%%%%%%%
\section{FUTURE WORK}

Currently we are fine tuning PROOF for a single cluster and are working
with some "early adopter sites" that have setup dedicated clusters.

Several smaller and larger additions and refinements of the system are
already foreseen. General infrastructure for dynamic startup of
the session will allow PROOF to co-exist with traditional batch system
as well as Grid based environments. This could be extended to dynamic
allocation and release of slaves during the session, e.g. based on the
data set to be processed.

We are also working at combining multiple
clusters at geographically separated sites which  will require a
hierarchy of master servers. Further integration of PROOF and ROOT
will drive the implementation of event lists and friend trees.
We are also working on {\tt TTree::Draw()} style functionality.

The Grid middle-ware is in wild development at the moment and we are
following closely these developments to make sure it will support
the features we need, especially facilities for interactive high
priority jobs and data services.

%%%%%%%%%%%%%%%%%%%%%%%%%%%%%%%%%%%%%%%%%%%%%%%%%%%%%%%%%%%%%%%%%%%%%%%%%
\section{CONCLUSIONS}

We demonstrated the PROOF system, part of the widely adopted ROOT
analysis environment, that allows users to harness the
power of a large cluster of workstations from their desktops.
PROOF greatly extends the amount and range of data that can be
interactively analyzed.

\begin{acknowledgments}
This material is based upon work supported by the National Science
Foundation under Grant No. 0219063.
\end{acknowledgments}

\end{document}